\documentstyle[12pt]{article}

\newcommand{\newsection}[1]
{\vspace{5mm}
\pagebreak[3]
\addtocounter{section}{1}
\setcounter{equation}{0}
\setcounter{subsection}{0}
\setcounter{footnote}{0}
\begin{flushleft}
{\large\bf \thesection. #1}
\end{flushleft}
\nopagebreak
\medskip
\nopagebreak}

\newlength{\extraspace}
\newlength{\extraspaces}
\def\appendix#1{\addtocounter{section}{1}\setcounter{equation}{0}
\renewcommand{\thesection}{\Alph{section}}
\section*{Appendix\thesection\protect\indent \parbox[t]{11.715cm} {#1}}
\addcontentsline{toc}{section}{Appendix \thesection\ \ \ #1} }
\newcommand{\complex}{{\bb C}} 
\newcommand{\zed}{{\bb Z}} 
\newcommand{\real}{{\bb R}} 
\newcommand{\zeds}{{\bbs Z}} 
\newcommand{\id}{{1\!\!1}} 

\newif\ifold             \oldtrue            \def\new{\oldfalse}
\def\arraymode{\ifold\relax\else\displaystyle\fi} 
\def\@arrayskip{\ifold\baselineskip\z@\lineskip\z@
     \else
     \baselineskip\minarrayskip\lineskip2\minarrayskip\fi}
\def\@arrayclassz{\ifcase \@lastchclass \@acolampacol \or
\@ampacol \or \or \or \@addamp \or
   \@acolampacol \or \@firstampfalse \@acol \fi
\edef\@preamble{\@preamble
  \ifcase \@chnum
     \hfil$\relax\arraymode\@sharp$\hfil
     \or $\relax\arraymode\@sharp$\hfil
     \or \hfil$\relax\arraymode\@sharp$\fi}}
\def\@array[#1]#2{\setbox\@arstrutbox=\hbox{\vrule
     height\arraystretch \ht\strutbox
     depth\arraystretch \dp\strutbox
     width\z@}\@mkpream{#2}\edef\@preamble{\halign \noexpand\@halignto
\bgroup \tabskip\z@ \@arstrut \@preamble \tabskip\z@ \cr}%
\let\@startpbox\@@startpbox \let\@endpbox\@@endpbox
  \if #1t\vtop \else \if#1b\vbox \else \vcenter \fi\fi
  \bgroup \let\par\relax
  \let\@sharp##\let\protect\relax
  \@arrayskip\@preamble}
\makeatother

\font\mybb=msbm10 at 12pt
\def\bb#1{\hbox{\mybb#1}}
\font\mybbs=msbm10 at 9pt
\def\bbs#1{\hbox{\mybbs#1}}

\def\nn{\nonumber}
\newcommand{\tr}[1]{\:{\rm tr}\,#1}
\newcommand{\Tr}[1]{\:{\rm Tr}\,#1}
\def\e{{\,\rm e}\,}

\newcommand{\non}{\nonumber \\*}

\def\be{\begin{equation}}
\def\ee{\end{equation}}
\def\bea{\begin{eqnarray}}
\def\eea{\end{eqnarray}}
\def\bd{\begin{displaymath}}
\def\ed{\end{displaymath}}

\def\dd{{\rm d}}

\newcommand{\beq}{\begin{eqnarray}}
\newcommand{\eeq}{\end{eqnarray}}

\begin{document}

\begin{flushright}

\baselineskip=12pt
HWM01--2\\ EMPG--01--01\\ hep--th/0101216\\
\hfill{ }\\
January 2001
\end{flushright}

\begin{center}

{\large\bf DISCRETE NONCOMMUTATIVE GAUGE THEORY\footnote{\baselineskip=12pt
Based on invited lecture given at the Euroconference ``Brane New World and
Noncommutative Geometry'', Villa Gualino, Torino, Italy, October 2--7 2000. To
be published in the proceedings by World Scientific.}}

\baselineskip=12pt

\vspace{.5cm}

{\bf Richard J.~Szabo}
\\[3mm]
{\it Department of Mathematics, Heriot-Watt University\\ Riccarton,
Edinburgh EH14 4AS, Scotland}\\  {\tt R.J.Szabo@ma.hw.ac.uk} \\[10mm]

\end{center}

\begin{center}
\begin{minipage}{15cm}
\small

\baselineskip=12pt

A review of the relationships between matrix models and noncommutative gauge
theory is presented. A lattice version of noncommutative Yang-Mills theory is
constructed and used to examine some generic properties of noncommutative
quantum field theory, such as UV/IR mixing and the appearence of
gauge-invariant open Wilson line operators. Morita equivalence in this class of
models is derived and used to establish the generic relation between
noncommutative gauge theory and twisted reduced models. Finite dimensional
representations of the quotient conditions for toroidal compactification of
matrix models are thereby exhibited. The coupling of noncommutative gauge
fields to fundamental matter fields is considered and a large mass expansion is
used to study properties of gauge-invariant observables. Morita equivalence
with fundamental matter is also presented and used to prove the equivalence
between the planar loop renormalizations in commutative and noncommutative
quantum chromodynamics.

\end{minipage}
\end{center}

\baselineskip=14pt

\newsection{Matrix Models and Noncommutative Gauge Theory}

In this article we will discuss the intimate relationship that exists between
Yang-Mills theory on a noncommutative space and large $N$ matrix models which
are conjectured to provide nonperturbative definitions of string theory and
M-Theory. This paper is based on the articles \cite{AMNS1}--\cite{AMNS3}. A
related review can be found in \cite{Mak}. We will begin by recalling how
noncommutative gauge theory first appeared within the context of
nonperturbative string theory.

\subsubsection*{The IKKT Matrix Model}

The IKKT matrix model \cite{IKKT} is defined as the dimensional reduction to a
point of ten-dimensional maximally supersymmetric Yang-Mills theory. The action
is
\beq
S_{\rm IKKT}=-\frac1{g^2}\,\Tr\left(\frac14\,\left[X^i\,,\,X^j\right]^2+
\frac12\,\overline{\Psi}~\gamma_i\,\Bigl[X^i\,,\,\Psi\Bigr]\right) \ ,
\label{SIKKT}\eeq
where $X^i$, $i=1,\dots,10$, are $N\times N$ Hermitian matrices, whose
eigenvalues represent the coordinates of a D-instanton in ten dimensional
spacetime, and $\Psi$ are $N\times N$ Hermitian matrices which are
Majorana-Weyl spinors in ten dimensions. The $X^i$'s are the reductions of the
ten dimensional gauge fields and $\Psi$ the reductions of their superpartners.
In the double scaling limit $N\to\infty$, $g^2\to0$ with the product $Ng^2$
finite, the model (\ref{SIKKT}) is conjectured to provide a nonperturbative
definition of Type IIB superstring theory \cite{IKKT}. Precisely, it is related
to the Green-Schwarz formulation of the Type IIB string in the Schild gauge. It
is also related to various other matrix models in string theory. For instance,
by compactifying one of the directions on a circle ${\bf S}^1$ one can recover
the BFSS matrix model \cite{BFSS} which is conjectured to provide a
nonperturbative definition of M-Theory. By compactifying two of the directions
on a torus ${\bf T}^2={\bf S}^1\times{\bf S}^1$, one can arrive at the DVV
matrix string theory which proposes a nonperturbative definition of Type IIA
superstring theory \cite{DVV}.

\subsubsection*{Toroidal Compactification}

The fact that the spacetime of the model (\ref{SIKKT}) is described by mutually
noncommuting matrices $X^i$ suggests that it should be related in some way to
noncommutative geometry \cite{Witten1}. This feature was first made precise in
\cite{CDS} in the following way. Let us compactify the coordinates $X^i$ on a
hypercubic $D$-dimensional torus ${\bf T}^D$ of sides $R^i$ for $i=1,\dots,D$.
Because of the $U(N)$ gauge symmetry of the action (\ref{SIKKT}), it should be
invariant under periodic shifts of the matrices $X^i$, $i=1,\dots,D$ around the
cycles of this torus up to unitary conjugation. Thus the toroidal
compactification of the IKKT matrix model is tantamount to finding unitary
matrices $U_i$ such that
\beq
X^i+2\pi R^i\,\delta^i_j\,\id_N=U_j^{-1}\,X^i\,U_j
\label{quotientcond}\eeq
for each $i,j=1,\dots,D$. Taking the trace of both sides of the equations
(\ref{quotientcond}) shows that they cannot be solved by finite dimensional
matrices. It is, however, straightforward to solve them by operators which act
on an infinite dimensional Hilbert space, i.e. by setting $N\to\infty$. By
applying the translations (\ref{quotientcond}) in two different directions it
is easily seen that the unitary operators $U_i$ must satisfy the consistency
conditions $[U_iU_jU_i^{-1}U_j^{-1},X^k]=0$ for all $i,j,k=1,\dots,D$, which
imply that the unitary operators $U_i\,U_j\,U_i^{-1}\,U_j^{-1}$ can be
represented via multiplication by some phases $\e^{-2\pi i\,\Theta^{ij}}$,
$\Theta^{ij}\in\real$, on this Hilbert space. The operators $U_i$ thereby obey
the commutation relations
\beq
U_i\,U_j=\e^{-2\pi i\,\Theta^{ij}}\,U_j\,U_i \ .
\label{NCtorus}\eeq
We will assume throughout that the antisymmetric $D\times D$ matrix
$\Theta^{ij}$ is invertible (so that $D=2d$ is even).

The algebraic relations (\ref{NCtorus}) are the defining presentation of the
noncommutative torus ${\bf T}_\Theta^D$ \cite{CDS}. The operators $U_i$
generate the algebra of smooth functions on ${\bf T}_\Theta^D$ through the
generalized Fourier series expansions
\beq
f=\sum_{\vec m\in\zeds^D}f_{\vec m}~U_1^{m_1}\cdots U_D^{m_D}
\label{fnUexp}\eeq
where $f_{\vec m}$ lives in an appropriate Schwartz space of sequences of
sufficiently rapid decrease. The $U_i$'s may be represented in terms of
Hermitian coordinate operators $\hat x^i$ as
\beq
U_i=\e^{2\pi i\,\hat x^i/R^i} \ , ~~ \left[\hat x^i\,,\,\hat x^j\right]=
\frac{i\,R^iR^j\,\Theta^{ij}}{2\pi} \ .
\label{Uhatxrep}\eeq
In the limit $\Theta\to0$ the expansion (\ref{fnUexp}) becomes the usual
Fourier mode expansion for functions on the ordinary torus ${\bf T}^D$.

\subsubsection*{The Moyal Product}

The algebra (\ref{NCtorus},\ref{fnUexp}) can be alternatively represented by
deforming the usual pointwise multiplication in the algebra $C^\infty({\bf
T}^D)$ of smooth functions on the ordinary torus to the associative, non-local
star-product
\bea
f(x)\star g(x)&=&f(x)~\exp\left(\frac i2\,\mbox{$\overleftarrow{\frac\partial
{\partial x^i}}$}\,\theta^{ij}\,\mbox{$\overrightarrow{\frac\partial
{\partial x^j}}$}\right)~g(x)\nn\\&=&f(x)\,g(x)+i\,\theta^{ij}\,\partial_i
f(x)\,\partial_jg(x)+{\cal O}\left(\theta^2\right)\nn\\&=&
\frac1{\pi^{D}|\det\theta|}\,\int\!\!\!\int\dd^Dy~\dd^Dz~f(x+y)
{}~g(x+z)~\e^{-2i(\theta^{-1})_{ij}\,y^iz^j} \ .
\label{starprodgen}\eea
On the torus ${\bf T}^D$, where $\theta^{ij}=R^iR^j\,\Theta^{ij}/2\pi$, we can
in addition represent this product by the Fourier series expansion
\beq
f(x)\star g(x)=\sum_{\vec m,\vec n\in\zeds^D}f_{\vec m}\,g_{\vec n-\vec m}
{}~\e^{-i\,\Theta^{ij}\,m_in_j}~\e^{i\,n_ix^i/R^i} \ .
\label{startorus}\eeq
In particular, via an integration by parts one can find that the natural trace
of products of functions in this deformed algebra coincides with that of the
undeformed algebra,
\beq
\Tr\,f\star g=\int\dd^Dx~f(x)\star g(x)=\int\dd^Dx~f(x)\,g(x) \ .
\label{tracestar}\eeq

\subsubsection*{Solving Quotient Conditions}

To solve the quotient conditions (\ref{quotientcond}) for toroidal
compactification of the IKKT matrix model, we introduce anti-Hermitian linear
derivations $\hat\partial_i$ which, together with the commutators of the
coordinate operators in (\ref{Uhatxrep}), obey the commutation relations
\beq
\left[\hat\partial_i\,,\,\hat x^j\right]=\delta^j_i \ , ~~
\left[\hat\partial_i\,,\,\hat\partial_j\right]=f_{ij} \ ,
\label{linderiv}\eeq
where $f_{ij}$ is an antisymmetric c-number tensor. Since then
$\hat\partial_iU_j=U_j(\hat\partial_i+2\pi i\delta_{ij}/R^i)$, these
derivations constitute a particular solution to the equations
(\ref{quotientcond}). A solution of the corresponding homogeneous equation can
be obtained from any function $A_i(\tilde U)$ of the generators $\tilde U_i$ of
the commutant of the algebra $C^\infty({\bf T}_\Theta^D)$, i.e. $[\tilde
U_i,U_j]=0~~\forall i,j$. They themselves generate a related noncommutative
$D$-torus. Therefore, the most general solution to the quotient conditions is
of the form
\beq
X^i=-i\left(R^i\right)^2\,\hat\partial_i+A_i\left(\tilde U\right) \ .
\label{gensoln}\eeq
 The infinite dimensional operators (\ref{gensoln}) represent a connection of a
gauge bundle (of topological charges $f_{ij}$) over the noncommutative torus.
When they are substituted back into the action (\ref{SIKKT}), one arrives at a
field theory which can be obtained from ordinary Yang-Mills gauge theory on
${\bf T}^D$ by replacing all products of fields by the Moyal product
(\ref{starprodgen}). This field theory is known as noncommutative gauge theory.

\newsection{Noncommutative Yang-Mills Theory}

The gauge theory obtained in the previous section is Yang-Mills theory on the
noncommutative torus which is defined by the action
\beq
S_{\rm NCYM}=-\frac1{g^2}\,\int\dd^Dx~F_{ij}(x)\star F^{ij}(x) \ ,
\label{NCYM}\eeq
where
\beq
F_{ij}=\partial_iA_j-\partial_jA_i-i\,A_i\star A_j+i\,A_j\star A_i \ .
\label{Fijdef}\eeq
This field theory possesses the noncommutative gauge symmetry
\beq
\delta_\lambda A_i=\partial_i\lambda+i\,\lambda\star A_i-i\,A_i
\star\lambda \ , ~~ \delta_\lambda F_{ij}=i\,\lambda\star F_{ij}
-i\,F_{ij}\star\lambda \ ,
\label{NCgaugesym}\eeq
where $\lambda\in C^\infty({\bf T}^D)$. Although the Moyal product leads to an
infinitely non-local interaction, this deformation of ordinary Yang-Mills
theory leads to a sensible quantum field theory. In fact, it is the unique
associative deformation which reduces to commutative Yang-Mills theory. It can
be analysed in perturbation theory by replacing the structure constants
$f_{abc}$ in the Feynman rules for ordinary non-abelian gauge theory everywhere
by the oscillatory momentum-dependent functions
$2i\sin(\frac12\,\theta^{ij}p_{ai}p_{bj})$, where $p_a$, $p_b$ and $p_c$ are
the incoming momenta of a three-gluon vertex. Many of the perturbative
properties of this theory can be thereby analysed.

While the effective noncommutative field theory so obtained has very natural
interpretations in both string theory and 11-dimensional supergravity, it has
required a stringent large $N$ limit to be taken in the IKKT matrix model and
the information originally encoded by the matrix dynamics has been lost. It
would be interesting to see if the origins of noncommutative geometry persists
at the level of a finite dimensional matrix model. This would enable, at least
in principal, the usage of standard numerical and analytical methods from
matrix model technology to solve a host of problems in string theory. In
\cite{AMNS1}--\cite{AMNS3} it was shown precisely how to do this via a lattice
version of noncommutative gauge theory. In addition to providing finite
dimensional versions of the above construction, it allows one to analyse
various properties of noncommutative field theory and also to explain why
noncommutative gauge theory arises so naturally from reduced models of
Yang-Mills theory. A key feature of this analysis is that it is carried out
within the framework of regulated quantum field theory and hence all results
thereby obtained are rigorous.

\newsection{Properties of Noncommutative Quantum Field Theory}

Let us first summarize some of the basic properties of noncommutative field
theories that will be analysed within the lattice approach to noncommutative
gauge theory in subsequent sections.

\subsubsection*{UV/IR Mixing}

There is a well-known distinction between planar and non-planar Feynman graphs
in noncommutative perturbation theory, analogous to that which arises in
multi-colour quantum chromodynamics. For illustration, consider massive
noncommutative $\phi^4$-theory in four dimensions, which is defined by the
interaction
\beq
\int\dd^4x~\phi(x)\star\phi(x)\star\phi(x)\star\phi(x)
=\prod_{a=1}^4\int\frac{\dd^4k_a}{(2\pi)^4}
{}~\phi(k_a)~\delta^{(4)}\left(\mbox{$\sum_bk_b$}\right)~V(k_1,k_2,k_3,k_4)
\label{phi4int}\eeq
with momentum space vertex function
\beq
V(k_1,k_2,k_3,k_4)=\prod_{a<b}\e^{-\frac i2\,k_{ai}\,\theta^{ij}\,k_{bj}} \ .
\label{phi4vertex}\eeq
The momentum dependent phase factor (\ref{phi4vertex}) depends on the cyclic
ordering of the vertex momenta $k_a$, and thereby contributes non-trivially
only to non-planar Feynman graphs. For instance, the one-loop planar and
non-planar contributions to the mass renormalization in this theory can be
written symbolically as (neglecting overall numerical factors)
\unitlength=1.00mm
\linethickness{0.4pt}
\begin{equation}\new{
\begin{picture}(60.00,10.00)
\thinlines
\put(0.00,0.00){\line(1,0){20.00}}
\put(10.00,4.20){\circle{8.00}}
\put(0.00,3.00){\makebox(0,0)[l]{$\scriptstyle{p}$}}
\put(15.00,4.00){\makebox(0,0)[l]{$\scriptstyle{k}$}}
\put(22.00,4.00){$~=~\int\frac{\dd^4k}{(2\pi)^4}~\frac1{k^2+\mu^2} \ , $}
\end{picture}}
\label{phi4planar}\end{equation}
\unitlength=1.00mm
\linethickness{0.4pt}
\begin{equation}\new{
\begin{picture}(60.00,10.00)
\thinlines
\put(10.00,4.00){\circle{8.00}}
\put(0.00,4.00){\line(1,0){12.00}}\put(16.00,4.00){\line(1,0){4.00}}
\put(0.00,7.00){\makebox(0,0)[l]{$\scriptstyle{p}$}}
\put(15.00,7.00){\makebox(0,0)[l]{$\scriptstyle{k}$}}
\put(22.00,4.00){$~=~\int\frac{\dd^4k}{(2\pi)^4}~\frac{\e^{ik_ip_j\theta^{ij}}}
{k^2+\mu^2} \ . $}
\end{picture}}
\label{phi4nonplanar}\end{equation}

{}From (\ref{phi4planar}) it is evident that the renormalizability properties
of planar noncommutative diagrams are the same as those of the corresponding
commutative theory (obtained by setting $\theta=0$) \cite{Filk}--\cite{MVS}.
This dispells the old belief that noncommutativity would generically serve as a
regulator of ultraviolet divergences in quantum field theory (at least for this
class of noncommutative geometries that arises naturally in string theory and
in M-Theory). On the other hand, the non-planar diagrams (\ref{phi4nonplanar})
exhibit the characteristic mixing of ultraviolet and infrared modes in
noncommutative perturbation theory. Namely, although for finite external
momentum $p$ the Feynman integral (\ref{phi4nonplanar}) is ultraviolet
convergent due to the oscillatory momentum-dependent phase factor, in the
infrared limit $p\to0$ the integral collapses to (\ref{phi4planar}) which is
ultraviolet divergent. The ultraviolet divergences in non-planar graphs have
been regulated by the noncommutativity, but these have reappeared as infrared
divergences. This is not completely surprising, since at $\theta=0$ these
divergences must reappear in some way, and this reemergence is characterized by
the low energy sector of the quantum field theory. It is a highly unexpected
result to have found infrared divergences in a massive quantum field theory.
Notice also that the $\theta\to0$ limit is not a smooth one \cite{MVS}. One of
the questions we will address in the following is whether this property is an
artifact of perturbation theory or if it manifests itself in nonperturbative
properties of the field theory.

This mixing of ultraviolet and infrared modes can in fact be argued for at a
heuristic level just from the existence of noncommutativity. Indeed, the
noncommutativity parameters lead to a correlation between the position
uncertainties in a given pair $i\neq j$ of spacetime directions of the form
$\Delta x^i=\theta^{ij}/\Delta x^j$. On the other hand, from the ordinary
Heisenberg uncertainty principle of quantum mechanics we have $\Delta
x^j=1/\Delta p^j$, and therefore the spatial extension of a particle in a
direction grows with its momentum in the transverse noncommutative directions.
The growth in the size of an object with its energy $E_i$ is characteristic of
string-modified uncertainty relations which have the form~\cite{Ven}
\beq
\Delta x^i=\frac1{E_i}+\alpha'\,E_i \ ,
\label{stringuncert}\eeq
and so the UV/IR mixing property of noncommutative quantum field theory can be
regarded as an intrinsically ``stringy'' feature of these models. This has been
at least part of the reason for the huge surge in activity in this models,
because as quantum field theories they are believed to capture many of the
generic properties of string theory.

\subsubsection*{Origins of Noncommutative Yang-Mills Theory}

Besides the manner outlined in section 1, there are two other ways that
noncommutative Yang-Mills theory arises as an effective field theory of string
dynamics. The first is as a description of the low-energy dynamics of open
strings on D-branes in background magnetic fields \cite{SW}. Formally, the
dynamics of the endpoints of the open strings is analogous to that described by
the quantum mechanical Landau Lagrangian
\beq
{\cal L}_{\rm L}=\frac{\mu^2}2\,\left(\dot x^i\right)^2+\frac B2\,
\epsilon_{ij}\,
x^i\dot x^j
\label{Landau}\eeq
which describes the motion of electrons in the plane $(x^1,x^2)$ in the
presence of a constant perpendicular magnetic field of strength $B$. In the low
energy limit $\mu\to0$, the coordinates become canonically conjugate operators
with commutator $[x^1,x^2]=i\,\theta$, where $\theta=1/B$. Thus the
configuration space of this simple model is deformed into a noncommutative
space by the presence of the magnetic field. In this same sort of low-energy
limit, the effective field theory governing the dynamics of D-branes in
magnetic fields is noncommutative gauge theory \cite{SW}.

The second appearence of noncommutative Yang-Mills theory in string theory
comes from expanding the spacetime variables $X^i$ of the IKKT matrix model
(\ref{SIKKT}) about a D-brane background, which is characterized by a
particular noncommuting configuration of the variables with constant curvature
\cite{AIIKKT}. Again this background can only be represented by
infinite-dimensional matrices. This derivation has been used to suggest that
twisted large $N$ reduced models may provide a concrete, non-perturbative
definition of noncommutative gauge theory. In the following we shall examine,
within a much more general framework, the reasons why this proposal appears to
be correct.

\subsubsection*{One-Loop Renormalization of Noncommutative Gauge Theory}

The quadratic part of the one-loop effective action in momentum space for
noncommutative quantum electrodynamics in four dimensions has been computed to
be \cite{Haya}
\bea
\Gamma_{\rm quad}&=&\frac14\,\int\frac{\dd^4k}{(2\pi)^4}~\left\{\frac1{g^2}-
\frac1{8\pi^2}\left(\frac{11}3-\frac23\,n_f\right)\ln\left(\frac{\Lambda^2}
{k^2}\right)\right.\non&&+\left.\frac{11}{24\pi^2}\,\ln\left(\frac1{k^2\left[
k_i(\theta^{ij})^2k_j\right]}\right)\right\}\,F^2 \ ,
\label{quadeffaction}\eea
where $\Lambda$ is an ultraviolet cutoff and $F$ is the noncommutative field
strength tensor (\ref{Fijdef}). The second term in (\ref{quadeffaction}) comes
from the planar one-loop Feynman diagrams while the third one is due to the
non-planar contributions. This result shows that, at the level of planar
noncommutative diagrams, i.e. in the limit $\theta\to\infty$, the
noncommutative $U(1)$ gauge theory has a running coupling constant which 
coincides with that of ordinary $SU(\infty)$ Yang-Mills theory.
In particular, ignoring issues related to the center $U(1)$ of the gauge group
when $N>1$ \cite{Adi}, we see that $U(N)$
noncommutative Yang-Mills theory with $n_f$ flavours of fundamental (fermion)
matter fields is equivalent to ordinary $U(\infty)$ Yang-Mills theory coupled
to $n_f\cdot N$ flavours of matter. As we will explain in the following, the
feature that noncommutative Yang-Mills theory at large $\theta$ coincides with
large $N$ commutative gauge theory is a consequence of Morita equivalence of
noncommutative gauge theories, which can also be thought of as a stringy
characteristic within the present class of quantum field theories.

\newsection{Lattice Regularization}

The questions raised thus far will be answered via a lattice regularization of
noncommutative field theory. Since, as we have discussed, constant
noncommutativity parameters do not cure a quantum field theory of its
ultraviolet divergences, this formalism will also provide a natural ultraviolet
regulator for these models. Our starting point will be the definition of a
noncommutative space of dimension $D$ which is defined by Hermitian coordinate
operators $\hat x^i$ that obey the commutation relations
\beq
\left[\hat x^i\,,\,\hat x^j\right]=i\,\theta^{ij}
\label{NCspacedef}\eeq
of noncommutative $\real^D$, where $i,j=1,\dots,D$. The lattice discretization
is defined by restricting the spacetime points of $\real^D$ to the discrete
values $x^i\in\epsilon\,\zed$, where $\epsilon$ is the lattice spacing. This
leads to a compact momentum space with periodically identified momenta $k_i\sim
k_i+\frac{2\pi}\epsilon\,\delta_{ij}$ for $j=1,\dots,D$. As a consequence, the
lattice discretization of noncommutative spacetime requires that the operators
$\hat x^i$ obey the constraint
\beq
\e^{i(k_i+\frac{2\pi}\epsilon\,\delta_{ij})\hat x^i}=\e^{ik_i\hat x^i}
\label{hatxconstr}\eeq
for each $j=1,\dots,D$. By multiplying both sides of (\ref{hatxconstr}) by the
operator $\e^{-ik_i\hat x^i}$ we learn two things. First of all, $\e^{2\pi
i\hat x^i/\epsilon}=\id$ for $i=1,\dots,D$. This is just the usual constraint
that arises in lattice field theory, stating that the spacetime discretization
must be compatible with the spectra of the position operators. However, because
of (\ref{NCspacedef}), the Baker-Campbell-Hausdorff formula produces a
non-trivial c-number phase factor which leads to the constraint
$\theta^{ij}\,k_j\in2\epsilon\,\zed$, which when combined with the periodicity
of momentum space implies that the numbers $\frac\pi{\epsilon^2}\,\theta^{ij}$
are integral for each $i,j=1,\dots,D$.

This means that the momentum space is also discrete,
$k_i=2\pi(\Sigma^{-1})_i^a\,m_a$, where $m_a\in\zed$, $a=1,\dots,D$, have the
periodicities $m_a\sim m_a+\frac1\epsilon\,\Sigma_a^i$ for $i=1,\dots,D$. Thus
the spacetime coordinates are effectively restricted to lie on a {\it periodic}
lattice
\beq
x^i\sim x^i+\Sigma^i_a \ , ~~ a=1,\dots,D \ ,
\label{perlattice}\eeq
where the $D\times D$ period matrix $\Sigma$ is defined by
\beq
M^{ia}\,\Sigma_a^j-M^{ja}\,\Sigma_a^i=\frac{2\pi}\epsilon\,\theta^{ij}
\label{perioddef}\eeq
for some integral $D\times D$ matrix $M$. We have therefore found that lattice
regularization of noncommutative field theory implies that spacetime is
necessarily compact. Notice that the continuum limit $\epsilon\to0$ does not
commute with the commutative limit $\theta^{ij}\to0$, since the former limit
restores the infinite (noncommutative) spacetime $\real^D$ while the latter
limit shrinks it to a point. This discrete compactification is therefore just
the UV/IR mixing phenomenon that we encountered perturbatively in the previous
section. Here we have found that it arises at a completely non-perturbative
level.

It is possible to take a continuum limit to a noncommutative torus in this case
by letting the matrix elements $M^{ia}\to\infty$ \cite{AMNS3}. In the general
case, however, we obtain a discrete representation of the noncommutative torus
characterized by the algebra
\beq
\hat Z_a\,\hat Z_b=\e^{-2\pi i\,\Theta^{ab}}\,\hat Z_b\,\hat Z_a \ ,
\label{hatZalg}\eeq
where we have defined the single-valued coordinate operators
\beq
\hat Z_a=\e^{2\pi i\,(\Sigma^{-1})^a_i\,\hat x^i} \ ,
\label{hatZadef}\eeq
and
\beq
\Theta^{ab}=2\pi\left(\Sigma^{-1}\right)_i^a\,\theta^{ij}\left(\Sigma^{-1}
\right)^b_j=\epsilon\left(\Sigma^{-1}\right)^a_i\,M^{ib}-
\epsilon\left(\Sigma^{-1}\right)^b_i\,M^{ia}
\label{Thetaabdef}\eeq
are the corresponding dimensionless noncommutativity parameters which are
rational-valued. Furthermore, on the lattice one should as usual consider not
the linear derivations $\hat\partial_i$ defined by (\ref{linderiv}), but rather
the shift operators $\hat D_i=\e^{\epsilon\,\hat\partial_i}$ which affect
translations in units of the lattice spacing $\epsilon$. They act on the
coordinate operators (\ref{hatZadef}) as
\beq
\hat D_i\,\hat Z_a\,\hat D_i^\dagger=\e^{2\pi i\,\epsilon\,(\Sigma^{-1})^a_i}
\,\hat Z_a \ .
\label{DZact}\eeq

To construct a field theory on the noncommutative lattice, we introduce a map
$\hat\Delta(x)$ which provides an isomorphism between the algebra of finite
dimensional Weyl operators $\hat f$ and the noncommutative algebra of lattice
fields $ f(x)$ with a discrete version of the Moyal product. If $ f_{\vec m}$
denotes the periodic lattice Fourier transform of the lattice field $ f(x)$,
then one can define its corresponding Weyl operator by the Fourier series
\beq
\hat f=\frac1{\left|\det\frac1\epsilon\,\Sigma\right|}\,
\sum_{\vec m}\e^{2\pi i\,(\Sigma^{-1})_i^a\,m_a\,\hat x^i}\, f_{\vec m} \ ,
\label{Weylopdef}\eeq
where the sum runs over all integral vectors $\vec m\in\zed^D$ modulo the
periodicity of momentum space. The expression (\ref{Weylopdef}) can be written
as~\cite{AMNS1,BarsMinic}
\beq
\hat f=\sum_x f(x)\,\hat\Delta(x) \ ,
\label{WeylopDelta}\eeq
where the sum runs over all lattice points $x\in\epsilon\,\zed^D$ modulo the
periodicity (\ref{perlattice}) and
\beq
\hat\Delta(x)=\frac1{\left|\det\frac1\epsilon\,\Sigma\right|}\,
\sum_{\vec m}\left[\,\prod_{a=1}^D\hat Z_a^{m_a}\,\prod_{b<a}
\e^{\pi i\,m_a\,\Theta^{ab}\,m_b}\right]~\e^{-2\pi i\,(\Sigma^{-1})_i^a
\,m_a\,x^i} \ .
\label{hatDeltadef}\eeq
Because of the identity
\beq
\frac1{\left|\det\frac1\epsilon\,\Sigma\right|}\,\sum_{\vec m}
\e^{-2\pi i\,(\Sigma^{-1})_i^a\,m_a\,x^i}=\delta_{x,0\,({\rm mod}\,\Sigma)} \ ,
\label{latticedelta}\eeq
it follows that at $\theta=0$ the map (\ref{hatDeltadef}) reduces trivially to
$\delta(x-\hat x)$. However, in the generic noncommutative case it defines a
very complicated transformation.

If Tr is any suitably normalized trace on the algebra of Weyl operators, then
it is straightforward to show that the trace of the map (\ref{hatDeltadef}) is
$\Tr\hat\Delta(x)=1$. Furthermore, these maps form a mutually orthonormal
system of operators on the lattice,
$\Tr\hat\Delta(x)\,\hat\Delta(y)=\delta_{xy}$. From these properties it follows
that the operator trace Tr may be uniquely represented in terms of a lattice
sum as
\beq
\Tr\hat f=\sum_x f(x) \ ,
\label{Trsum}\eeq
and one can construct the Wigner map
\beq
 f(x)=\Tr\left(\hat f\,\hat\Delta(x)\right)
\label{Wignerdef}\eeq
which is the inverse of the Weyl map (\ref{WeylopDelta}). In particular,
applying the Wigner map (\ref{Wignerdef}) to the product $\hat f\,\hat g$ of
two Weyl operators leads to the definition of the lattice star-product,
\beq
 f(x)\star g(x)\equiv\Tr\left(\hat f\,\hat g\,\hat\Delta(x)\right)
=\frac1{\left|\det\frac1\epsilon\,\Sigma\right|}\,\sum_{y,z}
\e^{-2i(\theta^{-1})_{ij}\,y^iz^j}\, f(x+y)\,g(x+z) \ ,
\label{latticestarprod}\eeq
where in the second equality we have assumed that $M^{-1}$ is an integral
matrix. The expression (\ref{latticestarprod}) is the lattice version of the
third equality in (\ref{starprodgen}).

\newsection{Discrete Noncommutative Yang-Mills Theory}

The construction of the previous section can be used to systematically
construct a lattice version of noncommutative gauge theory. As in ordinary
lattice gauge theory, the gauge fields are placed on links of the lattice,
which in the present case leads to the definition of the Weyl operators
\beq
\hat U_i=\sum_x\hat\Delta(x)\otimes U_i(x) \ ,
\label{hatUidef}\eeq
where $U_i(x)$ are $N\times N$ matrices. Unitarity of the Weyl operators
(\ref{hatUidef}), $\hat U_i^\dagger\,\hat U_i=\id$, is then equivalent to
star-unitarity of the corresponding lattice fields, $U_i(x)^\dagger\star
U_i(x)=\id_N$. A natural plaquette action may then be constructed from these
fields as
\bea
S_{\rm D}&=&-\frac1{g^2}\,\sum_{i\neq j}\Tr\tr_{(N)}\left[\hat U_i\left(
\hat D_i\,\hat U_j\,\hat D_i^\dagger\right)\left(\hat D_j\,\hat U_i^\dagger
\,\hat D_j^\dagger\right)\hat U_j^\dagger\right]\label{Sop}\\&=&-\frac1{g^2}\,
\sum_x\,\sum_{i\neq j}\tr_{(N)}\left[U_i(x)\star U_j(x+\epsilon\,\hat\imath)
\star U_i(x+\epsilon\,\hat\jmath)^\dagger\star U_j(x)^\dagger\right] \ ,
\label{Sdiscrete}\eea
where $\tr_{(N)}$ is the trace in the fundamental representation of the
$N\times N$ unitary group $U(N)$, and $\hat\imath$ is a unit vector in the
$i^{\rm th}$ direction of the lattice. The action (\ref{Sop},\ref{Sdiscrete})
is invariant under the noncommutative gauge transformations
\beq
\hat U_i~\longmapsto~\hat\omega\,\hat U_i\left(\hat D_i\,\hat\omega^\dagger
\,\hat D_i^\dagger\right) \ , ~~ U_i(x)~\longmapsto~\omega(x)\star U_i(x)
\star\omega(x+\epsilon\,\hat\imath)^\dagger \ ,
\label{latticestargauge}\eea
where the gauge operator $\hat\omega$ is unitary, or equivalently the $N\times
N$ gauge function $\omega(x)$ is star-unitary.

Note that this construction displays the manner in which the colour and
spacetime degrees of freedom are treated on equal footing in noncommutative
gauge theory. As we will see, this will be the essence of the relationship
between these models and large $N$ reduced models. The lattice regularization
now renders the quantum theory corresponding to the action (\ref{Sdiscrete})
rigorously defined. The lattice noncommutative gauge transformations
(\ref{latticestargauge}) form a finite-dimensional Lie group, and so the path
integral measure can be taken to be the corresponding Haar measure, i.e. that
which is invariant under left and right multiplications by elements of the
noncommutative gauge group, $\hat U_i\mapsto\hat\omega\,\hat U_i$ and $\hat
U_i\mapsto\hat U_i\,\hat\omega$. In the continuum limit $\epsilon\to0$, we
write $U_i(x)=\exp_\star i\,\epsilon\,A_i(x)$, where the star-exponential
$\exp_\star$ is defined by replacing ordinary products in the Taylor series
expansion of the exponential function by star-products. Then one can easily
work out the star-product of lattice gauge fields around a plaquette to be
\beq
U_i(x)\star U_j(x+\epsilon\,\hat\imath)
\star U_i(x+\epsilon\,\hat\jmath)^\dagger\star U_j(x)^\dagger=\exp_\star
i\,\epsilon^2\,F_{ij}(x) \ ,
\label{contlimit}\eeq
where $F_{ij}$ is the noncommutative field strength tensor (\ref{Fijdef}). Thus
in the continuum limit the action (\ref{Sdiscrete}) reduces to the usual
continuum noncommutative Yang-Mills action functional (\ref{NCYM}), and as such
it represents the natural noncommutative version of the Wilson plaquette action
for ordinary lattice gauge theory \cite{Wilson}.

Let us now describe the observables of lattice noncommutative gauge theory
\cite{AMNS1,AMNS2,AMNS3,IIKK}. In analogy with the commutative case, let us
choose an oriented contour ${\cal C}=\{i_1,i_2,\dots,i_L\}$ on the lattice
which is specified by $L$ links $i_a=\pm1,\pm2,\dots,\pm D$, $a=1,\dots,L$,
which start from the origin $x=0$ and end at the point
$\ell=\epsilon\,\sum_a\hat\imath_a$. We then introduce the noncommutative
version of the lattice parallel transport operator along the contour $\cal C$,
\beq
{\sf U}(x;{\cal C})=U_{i_1}(x)\star U_{i_2}(x+\epsilon\,\hat\imath_1)
\star\cdots\star U_{i_L}\left(x+\epsilon\,\sum_{a=1}^{L-1}
\hat\imath_a\right) \ ,
\label{partransport}\eeq
which transforms under noncommutative gauge transformations
(\ref{latticestargauge}) in the expected way as
\beq
{\sf U}(x;{\cal C})~\longmapsto~\omega(x)\star{\sf U}(x;{\cal C})
\star\omega(x+\ell)^\dagger \ .
\label{calUstargauge}\eeq
To construct star-gauge invariant operators from (\ref{partransport}), we note
that a simple application of the Baker-Campbell-Hausdorff formula gives
\bea
&&\e^{2\pi i(\Sigma^{-1})_i^a\,m_a\,x^i}\star\e^{2\pi i(\Sigma^{-1})_i^a\,n_a
\,x^i}\star\e^{-2\pi i(\Sigma^{-1})_i^a\,m_a\,x^i}\non&&~~~~~~=\e^{2\pi i
(\Sigma^{-1})_i^a\,m_a\,x^i}\star\e^{-2\pi i(\Sigma^{-1})_i^a\,m_a\,x^i}
\star\e^{2\pi i(\Sigma^{-1})_i^a\,n_a\,x^i}~
\e^{2\pi i\,n_a\Theta^{ab}m_b}\non&&~~~~~~=\e^{2\pi i(\Sigma^{-1})_i^a\,n_a
\bigl(x^i+2\pi\,\theta^{ij}(\Sigma^{-1})^b_j\,m_b\bigr)} \ ,
\label{expconj}\eea
which via Fourier transformation implies that plane waves in noncommutative
geometry affect the translations
\beq
\e^{2\pi i(\Sigma^{-1})_i^a\,m_a\,x^i}\star\omega(x)\star
\e^{-2\pi i(\Sigma^{-1})_i^a\,m_a\,x^i}=\omega(x+\ell_{\vec m})
\label{gaugetransl}\eeq
on arbitrary functions $\omega(x)$, where
\beq
\ell_{\vec m}^i=2\pi\,\theta^{ij}\left(\Sigma^{-1}
\right)^a_j\,m_a+\Sigma^i_a\,w^a
\label{videf}\eeq
with $w^a\in\zed$. This property, that unitary conjugation can induce spacetime
translations of a function, is particular to the noncommutative Moyal product.
It is related to the fact that noncommutative gauge theories are intimately
connected with general relativity, another feature of them which is tied to
their stringy nature.

{}From these properties it is straightforward to construct a gauge invariant
observable associated with the contour $\cal C$ as
\beq
{\sf O}({\cal C})=\sum_x\tr_{(N)}\,{\sf U}(x;{\cal C})\star
\e^{2\pi i(\Sigma^{-1})_i^a\,m_a\,x^i} \ .
\label{obsdef}\eeq
Its invariance follows from (\ref{calUstargauge}) and (\ref{gaugetransl}). The
integer vector $\vec m$ can be interpreted as the total momentum of the line
$\cal C$ and it is related to the separation of its endpoints by (\ref{videf}).
The most remarkable aspect of this construction is that we haven't had to
assume that the contour $\cal C$ is closed. Therefore, in marked contrast to
the commutative case, in noncommutative gauge theory there are gauge-invariant
observables associated with open contours. The price to pay for this extra
class of operators is that they are necessarily non-local, as the
noncommutative gauge symmetry is a geometrical one and so requires the usage of
the spacetime trace Tr in addition to the colour trace $\tr_{(N)}$ to define
star-gauge invariant operators. In the commutative limit $\theta=0$, we recover
the well-known fact that on a compact space the only gauge invariant
observables are the Polyakov lines which wind $w^a$ times around the compact
direction $a$. In that case, any function can be convoluted with the parallel
transport operator in (\ref{obsdef}) and we recover the usual local observables
of Yang-Mills theory. Notice also the feature that the size $\ell$ of the
contour grows with its momentum according to (\ref{videf}). This is just
another manifestation of the UV/IR mixing property that generically persists in
noncommutative field theories.

\newsection{Morita Equivalence}

We will now describe a remarkable symmetry on the space of noncommutative
(lattice) gauge theories. Consider {\it commutative} $U(N)$ lattice gauge
theory with 't~Hooft flux in dimension $D=2d$. The action is the usual Wilson
plaquette action~\cite{Wilson}
\beq
S_{\rm W}=-\frac1{g^2}\,\sum_x\,\sum_{i\neq j}\tr_{(N)}\left[U_i(x)\,
U_j(x+\epsilon\,\hat\imath)\,U_i(x+\epsilon\,\hat\jmath)^\dagger\,
U_j(x)^\dagger\right] \ ,
\label{Wilsonaction}\eeq
with the gauge fields obeying the twisted boundary conditions
\beq
U_i(x+\Sigma_j^a\,\hat\jmath)=\Omega_a(x)\,U_i(x)\,\Omega_a(x+\epsilon\,
\hat\imath)  \ .
\label{twistedbc}\eeq
Gauge fields which are multi-valued according to (\ref{twistedbc}) are only
defined on the universal covering space of the lattice because they carry a
non-vanishing flux. By writing the large gauge transformation (\ref{twistedbc})
along two different directions, we find that the transition functions
$\Omega_a(x)$ must satisfy the cocycle conditions
\beq
\Omega_a(x+\Sigma_i^b\,\hat\imath)\,\Omega_b(x)={\cal Z}_{ab}\,
\Omega_b(x+\Sigma_i^a\,\hat\imath)\,\Omega_a(x) \ .
\label{cocycle}\eeq
The phase factor ${\cal Z}_{ab}=\e^{2\pi i\,Q_{ab}/N}\in\zed_N$ is called the
``twist'', where $Q_{ab}$ is the integer 't~Hooft flux through the $(ab)$-th
two-cycle of the torus~\cite{tHooft}. If we choose $\Omega_a(x)=\Gamma_a$ to be
constant $SU(N)$ matrices, then (\ref{cocycle}) implies that the transition
functions obey the Weyl-'t~Hooft algebra
\beq
\Gamma_a\,\Gamma_b=\e^{2\pi i\,Q_{ab}/N}\,\Gamma_b\,\Gamma_a
\label{WeyltHooft}\eeq
which determines them as twist-eating solutions for $SU(N)$. Our task is to now
find the general form of the gauge fields which solve the twisted boundary
conditions (\ref{twistedbc}). For this, we map the problem onto an equivalent
one involving (commutative) Weyl operators. We introduce the map
(\ref{hatUidef}) and rewrite the action (\ref{Wilsonaction}) as in (\ref{Sop}).
The large gauge transformations (\ref{twistedbc}) may then be expressed in
terms of Weyl operators as
\beq
\left(\hat D_j\right)^{\Sigma_j^a/\epsilon}\,\hat U_i\,
\left(\hat D_j^\dagger\right)^{\Sigma_j^a/\epsilon}=\Gamma_a\,\hat U_i\,
\Gamma_a^\dagger \ .
\label{Weyltwistedbc}\eeq

\subsubsection*{Irreducible Representation of Twist Eaters}

Before solving (\ref{Weyltwistedbc}), we first need to digress a bit and
describe the representation theory of the Weyl-'t~Hooft algebra
(\ref{WeyltHooft}) \cite{twist}. For this, we use the discrete $SL(D,\zed)$
automorphism symmetry group of the torus to rotate $Q$ into a canonical
skew-diagonal form with skew-eigenvalues $q_\alpha\in\zed$, $\alpha=1,\dots,d$,
and define the three sets of $d$ integers
\beq
N_\alpha={\rm gcd}(q_\alpha,N) \ , ~~ \tilde N_\alpha=\frac N{N_\alpha} \ , ~~
\tilde q_\alpha=\frac{q_\alpha}{N_\alpha} \ .
\label{threeints}\eeq
A necessary and sufficient condition for the existence of twist-eating
solutions $\Gamma_a$ is that there exists an integer $\tilde N_0$ such that
$N=\tilde N_0\cdot\tilde N_1\cdots\tilde N_d$. The integer $\tilde
N_1\cdots\tilde N_d$ is then the dimension of the irreducible representation of
the Weyl-'t~Hooft algebra (\ref{WeyltHooft}), and the twist eaters $\Gamma_a$
may be defined on the $SU(N)$ subgroup $SU(\tilde N_1)\otimes\cdots\otimes
SU(\tilde N_d)\otimes SU(\tilde N_0)$ as
\bea
\Gamma_{2\alpha-1}&=&\id_{\tilde N_1}\otimes\cdots\otimes V_{\tilde N_\alpha}
\otimes\cdots\otimes\id_{\tilde N_d}\otimes\id_{\tilde N_0} \ , \non
\Gamma_{2\alpha}&=&\id_{\tilde N_1}\otimes\cdots\otimes\left(
W_{\tilde N_\alpha}\right)^{\tilde q_\alpha}\otimes\cdots\otimes\id_{\tilde
N_d}\otimes\id_{\tilde N_0} \ ,
\label{twistsoln}\eea
for $\alpha=1,\dots,d$. Here $(V_N)_{ab}=\delta_{a,b-1}$ and
$(W_N)_{ab}=\e^{2\pi i(a-1)/N}\,\delta_{ab}$ are the cyclic shift and clock
matrices of $SU(N)$ which obey the commutation relations
\beq
V_N\,W_N=\e^{2\pi i/N}\,W_N\,V_N \ .
\label{VpWpalg}\eeq
Note that the matrices (\ref{twistsoln}) commute with the $SU(\tilde N_0)$
subgroup of $SU(N)$ consisting of matrices of the form $\id_{\tilde
N_1}\otimes\cdots\otimes\id_{\tilde N_d}\otimes Z_0$, $Z_0\in SU(\tilde N_0)$.

By construction, for each $\alpha=1,\dots,d$ the integers $\tilde N_\alpha$ and
$\tilde q_\alpha$ are relatively prime, and so there exist integers $a_\alpha$
and $b_\alpha$ such that $a_\alpha\tilde N_\alpha+b_\alpha\tilde q_\alpha=1$.
We now introduce two diagonal $D\times D$ matrices $A$ and $\tilde{\cal N}$
whose diagonal elements are the integers $a_\alpha$ and $\tilde N_\alpha$,
respectively, each of which appear with multiplicity two. Likewise, we define
two skew-diagonal matrices $B$ and $\tilde Q$ whose skew-diagonal elements are
the integers $b_\alpha$ and $\tilde q_\alpha$. After using the $SL(D,\zed)$
automorphism group to rotate the flux matrix $Q$ back to general form, we find
that the resulting four integral $D\times D$ matrices $A$, $B$, $\tilde{\cal
N}$ and $\tilde Q$ are all mutually commutating and obey the matrix identity
\beq
A\tilde{\cal N}+B\tilde Q=\id_D \ .
\label{APBPrel}\eeq
As a consequence, these four matrices naturally produce a block matrix
\beq
\pmatrix{A&B\cr-\tilde Q&\tilde{\cal N}\cr}\in SO(D,D;\zed) \ .
\label{ABPQSODD}\eeq

\subsubsection*{Solving Twisted Boundary Conditions}

Using the above construction, it is now straightforward to write down the
general solution to the twisted boundary conditions (\ref{Weyltwistedbc}). The
details can be found in \cite{AMNS3}, and one finds
\beq
\hat U_i=\sum_{\vec m}\left[\,\prod_{a=1}^D\hat Z_a'^{\,m_a}\,
\prod_{b<a}\e^{\pi i\,m_a\,\Theta^{ab}\,m_b}\right]\otimes u_i(\vec m) \ ,
\label{twistgensoln}\eeq
where $u_i(\vec m)$ is an $\tilde N_0\times\tilde N_0$ matrix and the new
coordinate operators are
\beq
\hat Z_a'=\e^{2\pi i\,(\Sigma^{-1})_i^a\,\hat x^i}\otimes\prod_{b=1}^D\left(
\Gamma_b\right)^{B_{ab}} \ .
\label{Zaprime}\eeq
Because of the algebra (\ref{WeyltHooft}), the operators (\ref{Zaprime}) obey
the commutation relations (\ref{hatZalg}) with noncommutativity parameter
matrix
\beq
\Theta=-\tilde{\cal N}^{-1}\,B^\top \ \,
\label{ThetaMorita}\eeq
and also the relations (\ref{DZact}) with period matrix
$\Sigma'=\Sigma\,\tilde{\cal N}$. We see therefore that by absorbing the
't~Hooft flux $Q$ and some of the colour degrees of freedom into the coordinate
operators (\ref{Zaprime}), we have effectively arrived at a new toroidal
lattice which is now a noncommutative space.

We can now introduce a new Weyl-Wigner correspondence map $\hat\Delta'(x')$ by
substituting (\ref{Zaprime}) into (\ref{hatDeltadef}), where $x'$ are
coordinates on the new (noncommutative) torus. Using this new map, we may then
expand the operator $\hat U_i$ in terms of a new field $U_i'(x')$ which is a
single-valued $\tilde N_0\times\tilde N_0$ star-unitary matrix field on a
toroidal lattice with periods $\Sigma'$. The commutative Wilson plaquette
action (\ref{Wilsonaction}) then becomes the discrete noncommutative Yang-Mills
action (\ref{Sdiscrete}) of reduced gauge group rank $\tilde N_0$, with
rational-valued noncommutativity parameters (\ref{ThetaMorita}), and with new
Yang-Mills coupling constant $g'^{\,2}=Ng^2/\tilde N_0$. We have thereby found
that $U(N)$ commutative Yang-Mills theory with 't~Hooft flux is equivalent to a
noncommutative $U(\tilde N_0)$ gauge theory with single-valued gauge fields.
This transformation is an example of Morita equivalence. Note that it reduces
the rank of the gauge group and increases the size of the torus. In fact, when
$\tilde N_0=1$ (so that gauge group rank $N$ itself is the dimension of the
desired irreducible representation of the Weyl-'t~Hooft algebra), we see that
the colour degrees of freedom of any non-abelian gauge theory can be completely
absorbed into the noncommutativity of spacetime. This explains in part the
equivalences between ordinary and noncommutative quantum gauge theories that we
observed in section 3.

Morita equivalence is a standard equivalence relation among certain
noncommutative spaces. Here we have found that it is simply a result of a
change of basis $\hat\Delta(x)\leftrightarrow\hat\Delta'(x')$ for the mapping
between Weyl operators and lattice fields. This Morita equivalence is in fact
the noncommutative field theory version of the $SO(D,D;\zed)$ $T$-duality
symmetry of toroidally compactified string theory~\cite{Morita}. It is a
remarkable fact that this string theoretical duality is manifested explicitly
at a field theoretical level in noncommutative gauge theory. Furthermore, this
equivalence can be shown to hold also at the level of all quantum correlation
functions, and hence at the level of the full quantum field theory
\cite{AMNS3}. For example, under this duality the Polyakov lines of ordinary
gauge theory (with non-vanishing winding number $w^a$) on a torus are mapped
onto the open Wilson lines of the Morita equivalent noncommutative gauge theory
(with non-vanishing momentum $m_a$). Generally, at the level of observables,
the Morita equivalence acts precisely like a $T$-duality, in that it
interchanges momentum and winding modes $m_a\leftrightarrow w^a$.

\newsection{Twisted Eguchi-Kawai Model}

We are now ready to naturally describe the link between noncommutative gauge
theories and twisted reduced models. Let us reduce the Wilson action
(\ref{Wilsonaction}) to a single plaquette, leaving a one-site $U(N)$ lattice
gauge theory, i.e. $\Sigma=\epsilon\,\id_D$, with 't~Hooft flux. By reducing
the model to a point $x=0$, we can obtain the values of gauge fields at the
other three sites of the plaquette by using the twisted boundary conditions
(\ref{twistedbc}) to get
$U_i(\epsilon\,\delta^a_j\,\hat\jmath)=\Gamma_a\,U_i(0)\,\Gamma_a^\dagger$. The
action (\ref{Wilsonaction}) then reduces to
\beq
S_{\rm TEK}=-\frac1{g^2}\,\sum_{i\neq j}{\cal Z}_{ij}\,\tr_{(N)}
\left(V_i\,V_j\,V_i^\dagger\,V_j^\dagger\right) \ ,
\label{TEK}\eeq
where $V_i=U_i(0)\,\Gamma_i$ are $N\times N$ unitary matrices and the phase
factor ${\cal Z}_{ij}=\e^{2\pi i\,Q_{ij}/N}$ is the twist that appeared in
(\ref{cocycle}). This unitary matrix model is known as the (twisted)
Eguchi-Kawai model \cite{TEK}. Such reduced models of gauge theory where
originally introduced as matrix models whereby the spacetime dependence of
gauge fields is absorbed into colour degrees of freedom, necessitating a large
$N$ limit to be taken. The twist is required so that the reduced model is
equivalent to the 't~Hooft limit of large $N$ quantum field theory on the
continuum spacetime. This connection with noncommutative gauge theories is very
natural, since both sorts of models describe a mixing of colour and spacetime
degrees of freedom. In order to reproduce continuum field theories, the
Eguchi-Kawai model was always studied in the large $N$ limit. But now we have
unravelled a much deeper connection between the finite $N$ version of the
twisted Eguchi-Kawai model and noncommutative lattice gauge theory, i.e. the
two models are Morita equivalent. In the continuum limit, which requires
$N\to\infty$, we now see why the reason for the intimate relationship between
noncommutative Yang-Mills theory and reduced models -- it is the simplest
instance of Morita equivalence of noncommutative geometries. Note that the
twist factor is related to a background flux, consistent with the fact that
noncommutative field theory arises from string theory in background magnetic
fields. We emphasize that the connection just presented between matrix models
and noncommutative gauge theory is rigorous because it is derived from
regulated quantum field theory.

A particularly interesting feature of the reduced model is that it yields a
finite dimensional representation of the noncommutative torus. The algebraic
relations (\ref{hatZalg},\ref{DZact}) are satisfied by the $N\times N$ matrices
\beq
\hat Z_i=\prod_{j=1}^D\left(\Gamma_j\right)^{B_{ij}} \ , ~~
\hat D_i=\Gamma_i \ ,
\label{finiteNCrep}\eeq
with rational noncommutativity parameters (\ref{ThetaMorita}) and period matrix
$\Sigma=\epsilon\,\tilde{\cal N}$. The problem of representing a continuum
noncommutative torus using such a finite dimensional approximation is a little
subtle and technically involved. The detailed, rigorous construction can be
found in \cite{LLS1}. However, given the matrix model (\ref{TEK}), which is the
unitary version of the IKKT matrix model obtained by exponentiating the
Hermitian matrices of (\ref{SIKKT}) according to $V_i=\e^{i\,X^i/\epsilon}$, we
can now write down a finite-dimensional version of the quotient conditions for
toroidal compactification of the matrix model. Exponentiation of
(\ref{quotientcond}) leads to the conditions
\beq
\hat Z_j\,V_i\,\hat Z_j^\dagger=\e^{2\pi i\,\delta_{ij}\tilde N_i/N}\,V_i
\label{unitaryquotientcond}\eeq
which define the compactification of the twisted Eguchi-Kawai model (\ref{TEK})
on a rectangular torus ${\bf T}^D$ of sides $\epsilon\,\tilde N_i$,
$i=1,\dots,D$. Now taking the trace of both sides of the equation
(\ref{unitaryquotientcond}) only implies that $V_i$ is a traceless unitary
matrix, which is a condition obeyed by the twist-eating solutions $\Gamma_i$.
Moreover, the consistency condition generated by (\ref{unitaryquotientcond})
yields precisely the defining relations (\ref{hatZalg}) of the noncommutative
torus. Because all parameters involved are rational numbers, finite-dimensional
representations of all these quantities exist, and so it is possible to
maintain a finite $N$ version of the noncommutative geometry described in
section~1. A particular solution to the quotient conditions comes from the
shift operators $\hat D_i$ in (\ref{finiteNCrep}), which modulo gauge
transformations is the vacuum configuration of the matrix model (\ref{TEK}).
The general solution to (\ref{unitaryquotientcond}) is then given by
\beq
V_i=\hat\Lambda_i\,\Gamma_i \ ,
\label{finiteNgensoln}\eeq
where the unitary matrices $\hat\Lambda_i$ generate the commutant of the
noncommutative torus (\ref{hatZalg}), i.e. $[\hat\Lambda_i,\hat Z_j]=0$. The
unitary matrices (\ref{finiteNgensoln}) represent a finite-dimensional
approximation to a generic gauge connection on the noncommutative torus of
topological charges $Q_{ij}$ \cite{LLS1}. The general solutions of the
commutant condition can be expanded in the Weyl basis of $gl(N,\complex)$,
whose lattice Fourier transform provides the map between finite dimensional
matrices and lattice fields. The corresponding expansion coefficients $U_i(\vec
m)$ of  $\hat\Lambda_i$ are now interpreted as the Fourier coefficients of some
fields $U_i(x)$ defined on a periodic lattice. When (\ref{finiteNgensoln}) is
then substituted back into the Eguchi-Kawai action (\ref{TEK}), one arrives at
the noncommutative lattice Yang-Mills action (\ref{Sdiscrete}) for gauge group
$U(1)$. For more details of this construction, see \cite{AMNS1,Mak}. In this
way we have arrived at another derivation of the relationship between reduced
models and noncommutative gauge theory. The two approaches are equivalent,
because the commutant algebra generated by the matrices $\hat\Lambda_i$ in fact
defines a Morita equivalent noncommutative torus \cite{AMNS1,CDS}.

\newsection{Coupling to Fundamental Matter Fields}

We will now examine to what extent the observables of noncommutative gauge
theory can be regarded as fundamental. For this, we minimally couple the
noncommutative gauge theory (\ref{Sdiscrete}) to complex scalar fields
$\vec\phi(x)$ in the fundamental representation of the $U(N)$ gauge group. The
matter-coupled action is
\beq
S_{\rm m}=-\sum_{x,i}\vec\phi(x)^\dagger\star U_i(x)\star\vec\phi(x+
\epsilon\,\hat\imath)+\mu^2\,\sum_x\vec\phi(x)^\dagger\,
\vec\phi(x) \ ,
\label{Smatter}\eeq
and, together with (\ref{latticestargauge}), it is invariant under the
star-gauge transformations
\beq
\vec\phi(x)~\longmapsto~\omega(x)\star\vec\phi(x) \ , ~~
\vec\phi(x)^\dagger~\longmapsto~\vec\phi(x)^\dagger\star\omega(x)^\dagger \ .
\label{phistargauge}\eeq
The free scalar field propagator is
\beq
\Bigl\langle\phi_a(x)^*\,\phi_b(y)\Bigr\rangle_0=\frac1{\mu^2}\,\delta_{ab}\,
\delta_{xy} \ ,
\label{scalarprop}\eeq
where all averages in the following will denote those for a fixed gauge
background, i.e. the gauge fields are not integrated over.

A real technical advantage of the lattice formulation of noncommutative gauge
theory now comes into play. The scalar matter fields in the present lattice
field theory may be integrated out analytically using the standard large mass
expansion, in powers of $\frac1{\mu^2}$, of lattice gauge theory. For instance,
we can evaluate the effective action functional thereby induced for the gauge
fields, which is defined by the perturbation series
\beq
{\sf S}_{\rm eff}[U]=-\ln\left[\,\sum_{n=0}^\infty\frac1{n!}\,
\left\langle\left(\sum_{x,i}\vec\phi(x)^\dagger\star U_i(x)\star\vec\phi(x+
\epsilon\,\hat\imath)\right)^n\,\right\rangle_0\,\right] \ .
\label{SeffU}\eeq
The expectation values in (\ref{SeffU}) may be computed by using the standard
Wick expansion and the propagator (\ref{scalarprop}). The series (\ref{SeffU})
can thereby be reduced to a geometric sum over contours on the lattice of the
same type that appears in ordinary lattice gauge theory, and we have
\beq
{\sf S}_{\rm eff}[U]=\sum_{{\cal C}\,{\rm closed}}\frac{\mu^{-2L({\cal C})}}
{L({\cal C})}\,\sum_x\tr_{(N)}\,{\sf U}(x;{\cal C}) \ ,
\label{SeffUclosed}\eeq
where $L({\cal C})$ is the number of links in the contour $\cal C$. In this way
we recover the closed loop observables (\ref{obsdef}) (of momentum $\vec
m=\vec0\,$) of noncommutative Yang-Mills theory.

To see how the star-gauge invariant observables associated with open Wilson
lines arise, consider a two-point function of the form
\beq
{\sf G}[F]=\left\langle\sum_x\vec\phi(x)^\dagger\star\vec\phi(x)\star F(x)
\right\rangle \ ,
\label{2ptfn}\eeq
where $F(x)$ can be regarded as the wavefunction of the composite operator
$\vec\phi(x)^\dagger\star\vec\phi(x)$. By Fourier transforming the function
$F(x)$ to momentum space, we can write (\ref{2ptfn}) as
\beq
{\sf G}[F]=\sum_{\vec m}F_{\vec m}\,\left\langle\sum_x\vec\phi(x)^\dagger
\star\e^{2\pi i(\Sigma^{-1})_i^a\,m_a\,x^i}\star\vec\phi(x+\ell_{\vec m})\right
\rangle \ ,
\label{2ptfnmomspace}\eeq
where $\ell_{\vec m}^i$ is defined in (\ref{videf}) and we have used
(\ref{gaugetransl}). The matter correlators in (\ref{2ptfnmomspace}) are now
straightforward to evaluate and are again given geometrically by a standard
sum-over-paths representation. This enables us to write
\beq
{\sf G}[F]=\sum_{\vec m}F_{\vec m}\,\sum_{{\cal C}\,:\,\hat\imath_1+
\dots+\hat\imath_{L({\cal C})}=\ell_{\vec m}}\mu^{-2L({\cal C})}\,
\sum_x\tr_{(N)}\,{\sf U}(x;{\cal C})\star\e^{2\pi i(\Sigma^{-1})_i^a\,m_a\,
x^i} \ ,
\label{2ptfnopen}\eeq
and so we recover the open Wilson line observables (of arbitrary momentum $\vec
m$) of noncommutative gauge theory. Note that in the commutative limit
$\theta=0$, only closed contours $\cal C$ contribute to the sum in
(\ref{2ptfnopen}) and are thereby independent of the integer vector $\vec m$.
Then, the Fourier sum in (\ref{2ptfnopen}) can be done explicitly, reinstating
the wavefunction $F(x)$ and thereby recovering the closed Wilson line
observables of ordinary gauge theory. Thus the open Wilson line observables of
noncommutative gauge theory play just as fundamental a role as the closed ones
do in ordinary Yang-Mills theory. The present demonstration of this feature is
in fact identical to the way that Wilson loops were originally
discovered~\cite{Wilson}.

\subsubsection*{Morita Equivalence with Fundamental Matter}

Let us now consider {\it commutative} $U(N)$ gauge theory minimally coupled to
$N_f$ flavours of fundamental scalar fields $\Phi(x)_{a,\alpha}$, where the
index $a=1,\dots,N$ labels colour and $\alpha=1,\dots,N_f$ indexes flavour. If
we take $N_f=N$, then we may regard $\Phi(x)$ as an $N\times N$ complex matrix
field. The action obtained by summing $N_f$ commutative actions analogous to
(\ref{Smatter}) can then be written as
\beq
S_{\rm m}^{(N_f)}=-\sum_{x,i}\tr_{(N)}\,\Phi(x)^\dagger\,U_i(x)\,
\Phi(x+\epsilon\,\hat\imath)+\mu^2\,\sum_x\tr_{(N)}\,
\Phi(x)^\dagger\,\Phi(x) \ .
\label{SmNf}\eeq
It possesses the local left $U(N)$ colour symmetry analogous to
(\ref{phistargauge}), and also a global right $U(N)$ flavour symmetry
$\Phi(x)\mapsto\Phi(x)\,h$, $h\in U(N_f)$. We can thereby use the global
$SU(N_f)$ flavour symmetry to mimic the adjoint $U(N)$ representation for these
fundamental matter fields. In particular, as in (\ref{twistedbc}) we can impose
twisted boundary conditions on the fields $\Phi(x)$ in the form
$\Phi(x+\Sigma_i^a\,\hat\imath)=\Gamma_a\,\Phi(x)\,\Gamma_a^\dagger$, where
here $\Gamma_a$ represents a large gauge transformation while
$\Gamma_a^\dagger$ represents a rotation in flavour space. We may now repeat
the procedure described in section~6 of obtaining a Morita equivalent
matter-coupled noncommutative gauge theory. One finds generally that ordinary
$U(N)$ gauge theory coupled to $n_f\cdot N$ flavours of fundamental matter
fields is equivalent to $U(\tilde N_0)$ noncommutative gauge theory with
$n_f\cdot\tilde N_0$ flavours of fundamental matter. In the case $\tilde
N_0=1$, we have thereby unveiled a proper explanation for the equivalences of
commutative and noncommutative matter-coupled quantum gauge theories mentioned
at the end of section~3. We stress once more that these equivalences are
completely rigorous within the present setting, since they are obtained in a
regularized quantum field theory.

\subsection*{Acknowledgments}

The author would like to thank A. Armoni for comments on the manuscript.
He also thanks the organisors and participants of the
Euroconference for having provided a stimulating environment, and in
particular C.-S.~Chu, J.~Fr\"ohlich, A.~Schwarz and M.~Sheikh-Jabbari
for interesting discussions. This work is supported in part by an Advanced
Fellowship from the Particle Physics and Astronomy Research Council (U.K.).

\end{document}